\documentclass[pdflatex,sn-mathphys-num]{sn-jnl}
\usepackage{graphicx}%
\usepackage{multirow}%
\usepackage{amsmath,amssymb,amsfonts}%
\usepackage{amsthm}%
\usepackage{mathrsfs}%
\usepackage[title]{appendix}%
\usepackage{xcolor}%
\usepackage{textcomp}%
\usepackage{manyfoot}%
\usepackage{booktabs}%
\usepackage{algorithm}%
\usepackage{algorithmicx}%
\usepackage{algpseudocode}%
\usepackage{listings}%
\usepackage{hyperref}
\hypersetup{
    colorlinks = false,
    filecolor = gray,
    linkcolor = blue,
    linkbordercolor = red,
    citecolor = green,
    citebordercolor = purple,
    urlcolor = darkgreen,
    urlbordercolor = green
}

\theoremstyle{thmstyleone}

\theoremstyle{thmstyletwo}%

\theoremstyle{thmstylethree}%

\begin{document}

\title[First Radio Neutrino \& Cosmic Rays Astronomy Workshop in India]{First Radio Neutrino \& Cosmic Rays Astronomy Workshop in India}

\author{Mohammad Ful Hossain Seikh$^1$, Victor Grachev, Ilya Kravchenko, Mohammad Sajjad Athar, Dave Besson$^1$}\email{fulhossain@ku.edu, zedlam@ku.edu}

\affil{\orgdiv{$^1$Department of Physics \& Astronomy}, \orgname{KU}, \orgaddress{\street{1251 Wescoe Hall Drive}, \city{Lawrence}, \postcode{66045}, \state{KS}, \country{USA}}}

\abstract{We summarize the motivation for (as well as the presentations from) the April, 2024 workshop held in India and focused on radiowave techniques for cosmic ray and neutrino detection. }

\maketitle

Over the past quarter century, the detection of radio-frequency (RF) radiation at macroscopic wavelengths has rapidly advanced into a competitive technique for observing cosmic particles, particularly charged cosmic rays and neutrinos. The use of radio emissions from extensive air showers for cosmic-ray detection has proven to be a reliable method, now offering reconstruction accuracy for the primary observables (energy and direction) on par with more traditional approaches. In the case of neutrinos, radio detection in dense media stands out as the most promising strategy for achieving the vast detection volumes necessary to observe neutrinos at energies beyond the PeV scale, as initially revealed by IceCube. 

This maturation and coalescence of contemporary experimental efforts followed earlier, pioneering theoretical calculations of the long-wavelength, meter-scale electromagnetic emissions created by evolving particle showers, which gave rise to an initial generation of experimental initiatives demonstrating viability of the radio-frequency approach. Such efforts are based on Askaryan's realization, in the early 60's, that showers evolving either in the atmosphere or in dense media should develop a negative charge excess as pre-existing atomic electrons are Compton and Bhabha scatter into the evolving shower and/or the population of shower positrons depleted through annihilation with pre-existing atomic electrons; that negative charge excess radiates `coherently' at wavelengths greater than the transverse size of the shower.  

In 1965, a research team led by Jelley (including Porter and Weekes) detected radio emissions at 44 MHz that coincident with extensive air showers. Their investigation aimed to observe the Cherenkov-like radiation predicted by Askaryan, which was expected to result from the negative charge excess in such showers. The signals were captured using an array of 72 full-wave dipole antennas, operated in conjunction with three trays of Geiger counters. For the next several years, follow-up experiments attempted to develop the technique further. However, the extant waveform-recording hardware was unable to resolve shower characteristics at the level of detail required to characterize (large) shower-to-shower fluctuations, and the technique was abandoned until the maturation of digital data acquisition instrumentation in the late 90's. In parallel, detailed simulation-based calculations of the expected RF signal profiles, utilizing increasingly available high-performance computing, demonstrated excellent consistency with experimental data, and, by extension, excellent predictive power.

Although extensive air showers could, in principle, be detected anywhere on Earth, detection of showers evolving in dense media required identification of a large, radio-transparent target medium which also lent itself to straightforward deployment. Lab measurements indicating long RF attenuation lengths (km-scale) identified cold polar ice as an ideal target, although it presented obvious challenges in accessibility. A pilot effort (RAMAND, and based at the Institute of Nuclear Research in Moscow) at Station Vostok, Antarctica, collected the first background data, but was discontinued in the late 80's with the collapse of the USSR. That was followed by the RICE experiment at the South Pole, which parasitically made their first deployment of instrumentation in holes drilled for the AMANDA (and later IceCube) experiments in 1995. In the late 90's, the GLUE team followed through on Askaryan's originally recommended strategy of observing neutrinos interacting in the lunar regolith using a custom data acquisition and analysis strategy, with 2.2 GHz data collected by the Goldstone radio telescope in California.

Following those initial efforts, experiments using radio-frequency technology to detect both cosmic rays and also neutrinos have mushroomed. Relative to `standard' astroparticle physics instrumentation (photomultiplier tubes, scintillators, e.g.), construction of a front-end radio-frequency antenna can be quickly completed by an undergraduate in the space of a day; 200-MHz bandwidth digital oscilloscopes can be purchased from ebay.com for \$100/channel, so the price of coaxial cable, filters, and adapters, rather than digitizing electronics, can dominate the total cost of a new RF experiment. A pre-existing and mature cosmic ray experiment, based on conventional technology and techniques, offers a propitious platform for a new RF detector, which benefits enormously in background reduction by restricting data capture to already-identified EAS triggers. Such is the current situation in India, where the GRAPES-3 experiment, located in Ooty, has a well-established record of interesting and novel cosmic ray physics over the last 10--15 years; that prospect has provided much of the impetus for the workshop held at Aligarh Muslim University in April, 2024, and summarized in these Articles, as follows:

\subsection*{Ultrahigh energy cosmic rays and neutrino flux models (Muzio) \cite{Muzio2025}}
provides an overview of  models which predict the observable characteristics of cosmic rays (primarily protons and iron nuclei) and neutrinos arriving at Earth. Such models must specify the energy spectrum, and particle composition of some source (Active Galactic Nuclei, Gamma Ray Bursts, etc.), then evolve that spectrum through the near-source medium, incorporating known cross-sections and also some assumptions regarding the composition, and temperature, near the source. The spectral characteristics at Earth are obtained after convolving the particle-specific source spectra with additional modeling of the interactions with the intervening interstellar and intergalactic medium (magnetic fields and target particles), and also accounting for spectral redshift. The so-called Griesen-Zatsepin-Kuzmin (GZK) effect provides strong linkage between the ultra-high energy charged particle cosmic ray (UHECR) and ultra-high energy neutrino (UHEN) spectra: high-energy protons photoproduce pions (both charged and neutral) via interactions with the cosmic-microwave background: $p+\gamma_{CMB}\to\Delta\to N\pi$, with $N$ a nucleon. If the daughter pion is neutral, an ultra-high energy gamma-ray is produced (UHEG); if the daughter pion is charged, neutrinos are produced via pion decay. This one process therefore connects the UHECR, UHEN and UHEG fluxes via the cosmic microwave background - measurement of any one of the three quantitatively predicts the magnitude of the other two.  

\subsection*{Radio Detection of ultra-high-energy Cosmic-Ray Air Showers (Schr\"oder) \cite{Schroeder2025}} presents an overview, including an historical retrospective, of the radio technique -- the author has two decades of experience in the field, beginning with the `early' contemporary  experiments in Germany, continuing through the TUNKA-Rex experiment in Tunka Valley, Siberia (among the highest-return, per dollar, astroparticle physics initiatives) and the current radio extensions of both the Auger (Argentina) and IceCube (South Pole) experiments. Over that time period, the field has made remarkable advances, to the point where radio techniques now give shower maximum (and therefore particle species) information competitive with `standard' methods. The details of the radio emission mechanism are also outlined, including the energy-dependent fractional contributions from geomagnetic vs. Askaryan charge-excess processes. The article reviews twenty experiments, either completed, in-progress, or planned, illustrating the rapid expansion of the field, as a whole.  

\subsection*{Fast radio bursts through radio astronomy (Sharma) \cite{Sharma2025}}
This review article, titled "Fast radio bursts through radio astronomy," discusses Fast Radio Bursts (FRBs), which are brief but extremely bright radio transients lasting mere milliseconds. These bursts are detected across the entire sky at gigahertz frequencies and are thought to come from highly energetic, distant sources, making them a significant enigma in the universe (although the same processes which drive pulsar formation and energy emission are currently under study, in the context of FRBs) . The transient and luminous nature of FRBs necessitates advanced detection methods, high-resolution data analysis, and specialized instruments for their study. The paper highlights the crucial role of radio telescopes globally in the discovery, characterization, and analysis of FRBs, offering an overview of major observatories worldwide and their contributions to the growing catalog of FRB detections. Parenthetically, a bit more than a year after the time when this paper was presented in Aprl 2024, the CHIME experiment (based in Canada) released a large catalog of FRB's, more than doubling the available data sample and portending significant future advances in the field.

\subsection*{Cosmic ray energy and composition measurements with GRAPES-3 and other experiments (Varsi) \cite{Varsi2025}} Fahim Varsi's contribution to this conference is particularly relevant to the theme and objective of the workshop, reporting on the most recent results on cosmic ray energy spectra and composition measurements recently obtained from the GRAPES-3 experiment (which detects EAS primarily via the muon content), based in Ooty, India. As typical processes producing cosmic rays generically lead to power-law energy spectra (dE/dN$\sim E^{-\gamma}$), breaks in the power-law exponents $\gamma$ indicate the emergence of new physical processes. The presence of the `knee' in the UHECR spectrum, for example, at 4 PeV, corresponds to a hardening of the spectral index from $\gamma=2.6\to\gamma=3.1$. This paper also reports the very recent (2023) observation of an additional spectral break in the UHECR charged proton spectrum below the knee at an energy of 166 TeV. Also highlighted are deficiencies in the current models used to derive information on the primary that initiated an observed EAS from the ground-level particle (or radio, e.g.) measurements.

\subsection*{Charged current neutrino scattering from nucleons (Athar) \cite{AtharCC2025}} Charged current neutrino scattering from nucleons, particularly in the few GeV energy range relevant for neutrino oscillation experiments, involves various processes like quasielastic scattering, meson production ($\pi$, kaons, $\eta$), and deep inelastic scattering. Sajjad Athar and colleagues have extensively studied these processes, focusing on their contributions to the cross-section and their dependence on kinematic variables like \(Q^{2}\) and \(W\). This contribution is closely connected to:

\subsection*{50 years of Neutrino Physics at Aligarh Muslim University	(Athar) \cite{Athar50Y2025}} details the long, half-century history of theoretical interpretation and calculation of neutrino cross-sections, dating back to the earliest fixed-target data from bubble-chamber experiments. Of particular relevance to this workshop, the authors have also made calculations of the atmospheric neutrino flux resulting from either decays of pions from the hadronic component of the shower, or semileptonic decays of charmed particles produced in the evolving shower. 

\subsection*{Cosmic Ray Sources and Detectors (Mohanty) \cite{Mohanty2025}} Prof. Mohanty (PI of the GRAPES-3 experiment) presents a high-level overview of cosmic ray measurements, beginning with Victor Hess' seminal measurement of an increase in the charged particle flux (as measured with a simple electroscope) with altitude in an ascending balloon. At low energies, space-based detectors measure primary cosmic ray particles; the limited size of such detectors, however, limits their sensitivity to 100 TeV. At higher energies, primary properties must be reconstructed from ground-level extensive air showers, such as those measured by the GRAPES experiment, at an elevation of 2240 m. Prof. Mohanty also reviews the interpretation of spectral data vis-a-vis extant models for UHECR production.

\subsection*{Neutrino sources: From meV to EeV	(Agarwalla) \cite{Agarwalla2025}} The neutrino energy spectrum, much like the electromagnetic spectrum, spans the range from primordial particle production (Cosmic Microwave Background photons, resulting from the Universe temperature cooling below 13.6 eV, or neutrinos [CMBN], resulting from the Universe temperature cooling below the weak scale) up to EeV energies ($>$20 orders of magnitude). The much-lower neutrino cross-section has two important implications, relative to photons: the lowest-energy neutrinos are conceded to be immeasurably feeble (in contrast to the CMB) while, at the highest-energy, the larger electromagnetic cross-section corresponds to mean free paths too small for UHEGR (ultra-high energy gamma rays) being observed at Earth, which has focused attention on UHEN for measurements of distant sources. This contribution reviews the myriad processes responsible for neutrino production, ranging from CMBN, through neutrinos produced by decay of radioactive elements in the Earth's crust (geoneutrinos), to ubiquitous solar neutrinos, to those produced beyond our solar system, and even our local galaxy. The energy scale accessible to individual neutrino detectors is typically dictated by the total instrumented size, and the spacing between individual sub-detectors, limiting sensitivity to typically 2--4 decades of neutrino energy. This review summarizes the processes responsible for producing observable neutrinos and also the experiments capable of measuring them.

\subsection*{Demonstrating the ability of IceCube DeepCore to probe Earth’s interior with atmospheric neutrino oscillations (Krishnamoorthi) \cite{Krishnamoorthi2025}}
The science output of the IceCube experiment spans glaciology, astroparticle and also (more mainstream) particle physics. The arrival direction of neutrinos measured in the 86-string photomultiplier tube array encodes information from individual sources (in the case of `hot spots', the Galactic Center, or the 2017 TXS blazar) - the quadratic variation of the angular source spot size varies with the one-dimensional angular resolution underscores the importance of excellent PMT timing. For the diffuse flux (for which either the source direction has been scrambled by interstellar and intergalactic magnetic fields, or the neutrinos are produced as secondary particles produced in the evolution of Extensive Air Showers), the sources are themselves isotropically distributed on the sky. It is also well-known that, via the MSW-effect, matter-induced oscillations of neutrinos traversing solid media result in a mix of the three neutrino generations ($\nu_e$, $\nu_\mu$ and $\nu_\tau$) differing from the fractional composition entering the medium. Quantitatively, the evolution of the population fractions are described by the neutrino `mixing' matrix, in direct analogy with the well-known quark `mixing' matrix. Since different neutrino (produced by decays of secondaries produced in UHECR-induced EAS) species produce different signatures as they traverse the ice, and leading to the IceCube detector, the observed neutrino mixture can be used to infer more detailed information about oscillations, as outlined in this contribution.

\subsection*{Askaryan Radio Detectors: the path towards IceCube-Gen2 Radio (Toscano) \cite{Toscano2025}}
Theoretical and experimental work in neutrino detection, using long-wavelength radio techniques, have been ongoing for well over the last half-century. Although progress has been steady, thus far there have been no positive UHEN detections. Three ingredients are critical for an experiment that (we can confidently predict) will have the sensitivity to not only detect neutrinos, but also perform true `neutrino astronomy' by mapping source directions of statistically significant samples (note that, at these high energies, the atmospheric neutrino background is negligible): i) low energy threshold with good background rejection (now established by the phased array triggering approaches incorporated into the ARA and RNO-G projects), ii) high radio-frequency transparency, with radio-frequency attenuation lengths of order 1 km or higher (now established at both Summit Station and also South Pole), and iii) large instrumented volume, to compensate for the miniscule neutrino fluxes at UHEN energies. Dr. Toscano outlines how, after years of development, the first two criteria are now satisfied; what remains is a large-scale drilling and deployment campaign, with 150 stations deployed to a depth of 100 meters, covering 100 square km aereally, to realize the last requirement.

\subsection*{Askaryan Radio Array (ARA): searching for the highest energy neutrinos (Seikh) \cite{SeikhARA2025}} Mr. Seikh is one of the key individuals currently conducting the search for UHEN using data from the 5 stations that comprise the ARA experiment at the South Pole. The analysis strategy, detailed herein, follows the same recipe that has been used since the initial search by the RICE experiment in 2000: 1) `clean' the data by identifying, and removing continuous wave contamination (Askaryan signals should be broadband in frequency and therefore impulsive [ns-scale] in the time domain), 2) identify and remove periods of high trigger rates, as well as periodic triggers, 3) via reconstruction, identify and remove both known RF sources (embedded radio-frequency calibration pulser transmitters in-ice, and also any `hotspots' emanating from the surface), 4) identify and remove non-impulsive sources, 5) suppress thermal fluctuations by requiring that all channels exhibit similar candidate signal characteristics with sufficiently high Signal-to-Noise (SNR, and, finally, 6) restricting the remaining candidate source to those that are `up-coming'. In this contribution, Mr. Seikh qualitatively and quantitatively outlines the path to isolating neutrino candidates with the ARA experiment.  

\subsection*{Physics of radio antennas (Seikh) \cite{SeikhAntennas2025}} Mr. Seikh's analysis of ARA data has been facilitated by his post-graduate degree in Electrical Engineering, with an emphasis on antenna design and calibration, with long hours spent in the University of Kansas anechoic chamber. Sharing many characteristics of RLC resonators, albeit with the capability to either transmit or reflect incident signal, RF antennas can be described using language familiar to the EE community. This contribution presents a detailed pedagogy on antenna properties and characteristics; unlike other expositions on the subject, the narrative here is presented from the perspective of an astroparticle physicist.

\subsection*{Backgrounds in UHE Cosmic Rays and Neutrino Detectors (Mikhailova) \cite{Mikhailova2025}}
As UHEN experiments have advanced, our understanding of the myriad backgrounds that may contaminate a search for neutrinos (or down-coming Askaryan signals resulting from EAS shower cores impacting the ice surface) has also evolved, in concert. In this contribution, Dr. Mikhailova focuses on two particular backgrounds that have recently come to light. The `triboelectric effect' results from electric discharges following wind-induced snow surface charge separation. Compiling and analyzing data from several UHEN experiments indicates that, at wind velocities exceeding 8--10 m/s (comprising typically a few percent of the livetime), trigger rates being to elevate, roughly linearly with wind speed, and saturating data acquisition (DAQ) throughput at wind velocities twice the threshold value. The second impulsive background source considered is RF emissions resulting from solar flares. Although previously observed by the RICE and ARA experiments, RNO-G has far exceeded those previous data analyses, in both statistics and quality. Perhaps most impressively, as described by Mihailova, since the source location in the sky is known, interferometric source reconstruction can be used to constrain the receiver antenna positions, in the ice, with a precision of 10 cm (and thereby approaching the precision on the phase center of the receiver antennas). The large sample of accessible solar flares permits an over-constrained geometry estimate, and also affords multiple internal cross-checks.

\subsection*{Askaryan and geomagnetic radiation from particle cascades (de Vries) \cite{deVries2025}}
provides a theoretical review of two fundamental mechanisms behind radio emission from particle cascades induced by cosmic rays and neutrinos; {\it Askaryan radiation}: radio waves produced by coherent charge asymmetry within particle a shower (in air or in a dielectric medium) and {\it geomagnetic radiation}: radio emission resulting from deflecting charged particles in Earth's magnetic field (in air). Prof. de Vries constructs an intuitive derivation of both mechanisms using the Lienard–Wiechert potentials. This involves starting from macroscopic current and charge distributions and working within the point-like approximation to derive the resulting radio emission expressions. 

\subsection*{Radio-frequency ice permittivity and the impact on neutrino detection experiments (Besson) \cite{Besson2025}} shows the importance of precise understanding of the radio-frequency response of ice, to the experiments seeking UHEN detection.

\subsection*{Outlook for the future}
There is clearly sufficient intellectual capital and will to propel future development of radio CR and neutrino detection within India. The relative economy of the necessary hardware makes such a prospect particularly attractive. As of this writing, interest seems to be coalescing around the prospect of a low-cost radio array co-deployed with the well-demonstrated and venerable GRAPES-3 in Ooty. The major challenge for such a co-deployment is the apparent incompatibility in energy scales -- GRAPES-3 sensitivity degrades above 10 PeV cosmic ray energies, whereas radio detection of CR's typically `turns on' at higher energies. Implementation of phased array techniques, already demonstrated elsewhere, can compensate and achieve an improvement in SNR$\propto\sqrt{N}$, where N is the number of receiver antennas participating in the event trigger. Most obviously, with event buffering, event triggers can be directly taken from GRAPES-3 and circumvent the self-triggering problem, provided the event latency is not prohibitively long. In any case, there is an opportunity for India to significantly advance it's national CR-detection program; hopefully this will be realized in the near future.

\end{document}